\documentclass[preprint,12pt]{elsarticle}

\usepackage{amssymb}
\usepackage{amsmath}

\usepackage{amsthm}
\usepackage{tikz}
\usepackage{graphicx}
\usepackage{nomencl}
\usepackage{subfigure}
\usepackage{xcolor}
\usepackage{color, soul}

\makenomenclature

\journal{International Journal of Heat and Mass Transfer}

\begin{document}

\begin{frontmatter}

\title{{Numerical Simulation of Turbulent Concentric Annular Pipe Flow using One-Dimensional Turbulence (ODT): Part 2: Heat Transfer}}


\author[add1,add2]{Pei-Yun Tsai}
\affiliation[add1]{
            addressline={Chair of Numerical Fluid and Gas Dynamics, Brandenburg University of Technology Cottbus-Senftenberg, Siemens-Halske-Ring 15a, 03046 Cottbus, Germany}
            }

\affiliation[add2]{
            organization={Scientific Computing (SC) Lab, Energy Innovation Center (EIZ), Brandenburg University of Technology Cottbus-Senftenberg},
            addressline={Siemens-Halske-Ring 15a, 03046 Cottbus, Germany}
            }

\author[add1,add2]{Heiko Schmidt}

\author[add1,add2]{Marten Klein}

\begin{abstract}
Turbulent concentric coaxial (annular) pipe flow with passive heat transfer is theoretically analyzed and numerically modeled in an extended parametric range using the stochastic one-dimensional turbulence (ODT) model. ODT provides predictive capabilities by fully resolving viscous, conductive, and turbulent advective transport processes along a representative radial coordinate within a dimensionally reduced and stochastic model formulation. Using a fixed model calibration for moderate and low Prandtl numbers, $Pr=0.71$ and $0.025$, effects of radius ratio, $\eta=R_{\rm i}/R_{\rm o}$ are investigated up to a highly turbulent flow regime. The analytical expression of the inner wall boundary layer yields a logarithmic law of the wall for the passive temperature. These results suggest that a conventional linear expression is inadequate for representing near-wall low-order statistics in the radial gap, in particular at the cylindrical inner wall. Additionally, the log-law region falls short if curvature and finite Reynolds number effects are not considered. Analytical boundary layer profiles fitting numerical predictions form the basis for heat transfer scaling relations. Heat transfer scalings are parameterized by a Nusselt correlation, which is extended to account for radius ratio effects. The findings demonstrate that the radius ratio has a significant impact on the thermal statistics over the two curved walls and should be considered even at high Reynolds numbers and low Prandtl numbers.
\end{abstract}

\begin{graphicalabstract}
\includegraphics[width=1.0\textwidth]{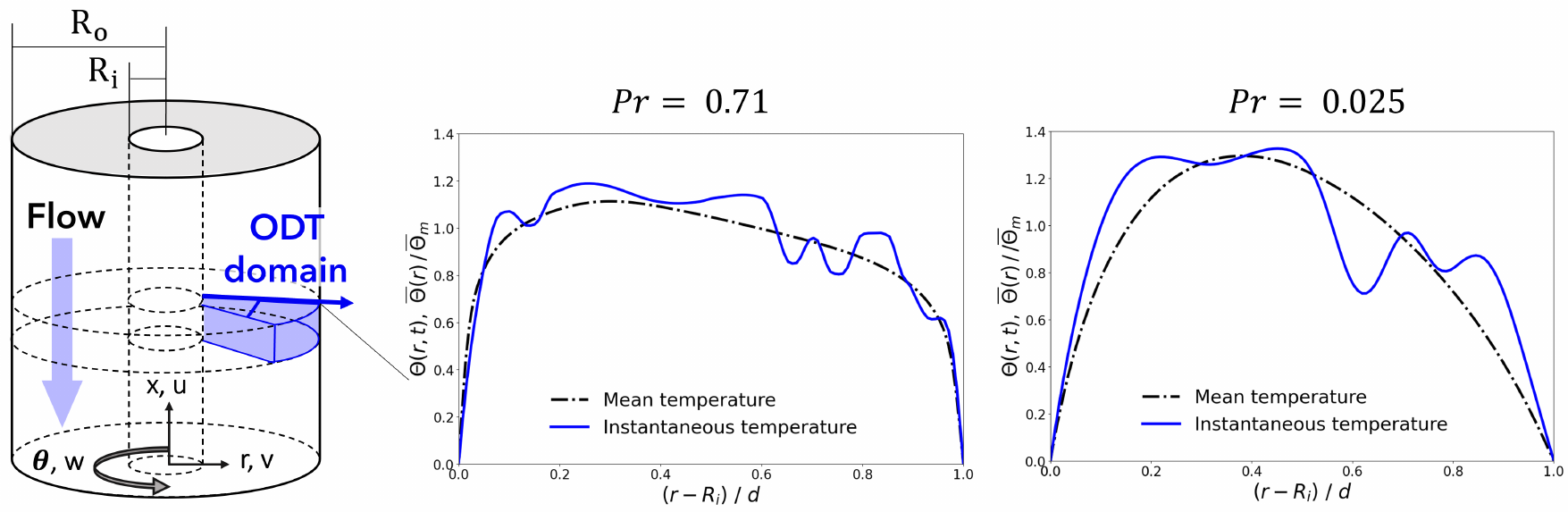}
\end{graphicalabstract}

\begin{highlights}
\item Prediction of heat transfer scalings in annular pipe flows by stochastic modeling
\item Systematic comparison of the effects of Prandtl number and Reynolds number
\item An analytical expression of the boundary layer with curvature effects is given
\item A Nusselt correlation is provided, extended to arbitrary radius ratios (gap width)
\end{highlights}


\end{frontmatter}



\section{Introduction}\label{sec:introduction}
A comprehensive understanding of the heat transfer and thermal properties of turbulent annular pipe flows under various conditions is crucial for predictive modeling and simulation of tubular heat exchangers and their operation.
Examples include geothermal heat exchangers, formulation of simplified heat exchanger models~\cite{raymond2015} for simulations of thermal energy storage systems~\cite{renaud2021} and high-temperature heat pumps~\cite{chen-tomac2023}, modeling of heat transfer in electrostatic precipitators in gas cleaning applications~\cite{shapiro1989,mendez2022}, among others.
Numerous studies have experimentally and numerically investigated heat transfer in turbulent annular pipe flows as canonical problem representative of the range of applications. Key areas of focus include the impact of spanwise curvature between the inner and outer cylindrical walls on the asymmetry of statistical moments in velocity and temperature fields~\cite{chung2002,chung2003,bagheri2020,bagheri2021}, turbulent structures~\cite{ishida2016friction,ishida2017turbulent}, and the Nusselt number~\cite{kays1963,quarmby1970turbulent,gnielinski2009,gnielinski2015}. Furthermore, research has explored the dependence of the ratio of heat flux~\cite{ould2009} on the two cylindrical walls, the Prandtl number~\cite{ould2010}, and the effects of boundary conditions~\cite{fukuda2020}. 

The early framework for annular heat transfer established basic solutions for varying boundary conditions. Kays and Leung (1963)~\cite{kays1963} provided the most reliable early numerical results of broad scope, deriving an asymptotic solution for the Nusselt number with arbitrarily prescribed heat flux. Their solutions covered a wide range of radius ratios and Prandtl numbers, validated by experimental data for air, Prandtl number $Pr=0.7$. Quarmby and Anand (1970)~\cite{quarmby1970turbulent} analyzed the heat transfer in concentric annuli with isothermal boundary conditions. They found that the constant wall temperature generally results in a lower Nusselt number than the uniform heat flux condition. Specifically, they demonstrated that the radius ratio effect on the Nusselt number is lower when the outer wall is heated than when the inner wall is heated. Yu et al. (2003,2005)~\cite{yu2003part1,yu2005part2,yu2005part3,yu2005part4} challenged foundational models, such as those by Kays and Leung, for their reliance on eddy viscosity, which they argued is fundamentally unsound for asymmetric flows like those in an annulus. They demonstrated that in these geometries, the radial location of maximum velocity does not coincide with the location of zero total shear stress. This displacement causes the eddy viscosity (the ratio of stress to velocity gradient) to become singular at the velocity peak and negative in the adjacent region. By correlating dimensionless turbulent shear stress directly, the authors provided more accurate Nusselt number predictions across diverse heating conditions.

Several direct numerical simulation (DNS) studies expanded the understanding of near-wall thermal structures by resolving small-scale dynamics without closures. Chung and Sung (2003)~\cite{chung2003} conducted a comprehensive DNS study on this topic, considering uniform heating of both cylindrical walls. The temperature was treated as a passive scalar, and simulations were tested for two radius ratios $\eta=0.1$ and $\eta=0.5$, at a bulk Reynolds number $Re_{D_{\rm h}}=8900$ and Prandtl number $Pr=0.71$. The radius ratio is defined as $\eta=R_{\rm i}/R_{\rm o}$, where $R_{\rm i}$ and $R_{\rm o}$ denote the radius of the inner and outer pipes, respectively. The bulk Reynolds number is defined as $Re_{D_{\rm h}}=\overline{u}_{\rm b} D_{\rm h}/\nu$, where $\overline{u}_{\rm b}$ represents the bulk mean velocity, $D_{\rm h}=2(R_{\rm o}-R_{\rm i})$ is the hydraulic diameter and $\nu$ denotes the kinematic viscosity of the fluid. Chung and Sung found that turbulent thermal flow structures are more active near the outer wall than near the inner wall. Ould-Rouiss \textit{et al.} (2009,2010)~\cite{ould2009,ould2010} investigated the influence of thermal boundary conditions on turbulent heat transfer in low-$Pr$ annular pipe flows by varying the heat flux ratio using DNS. Fukuda and Tsukahara (2020)~\cite{fukuda2020} used DNS to investigate the subcritical transition regime, where the friction Reynolds number $Re_{\tau}$ ranging from $50$ to $150$ and is equivalent to hydraulic diameter Reynolds numbers $Re_{D_{\rm h}}$ approximately between $2900$ and $9300$, and found that non-turbulent longitudinal-vortex clusters promote heat transfer even in intermittent states, reaching rates comparable to those in fully turbulent flow. They also examined the dependence of heat transfer on thermal boundary conditions, comparing uniform heat-flux (UHF) conditions with constant temperature difference (CTD). Bagheri and Wang (2021)~\cite{bagheri2021} carried out a systematic DNS study covering radius ratio ranging from $\eta=0.1$ to $\eta=0.7$ and analyzed the impact of curvature on the thermal properties as well as the interaction of thermal boundary layers formed along the two cylindrical walls at Reynolds number $Re_{D_{\rm h}}=1{,}77\times 10^4$ and Prandtl number $Pr=0.71$. More recently, Wang \textit{et al.} (2025)~\cite{wang2025thermal} extended the exploration to extreme curvature $\eta=0.01$ using sensitized Reynolds stress modeling (IIS-RSM). They found that flow and thermal properties depart dramatically from equilibrium (log-law) near the inner wall, resulting in suppressed turbulence production and weakened anisotropy.

Available experimental studies and advanced turbulence models have enabled the development of generalized correlations that connect high-fidelity simulations with practical engineering design. Gnielinski (2009, 2015)~\cite{gnielinski2009,gnielinski2015} summarized existing experimental data from the literature and proposed improved empirical Nusselt correlations for the dimensionless convective heat transfer coefficient, incorporating the radius ratio as a key parameter for cases where either the inner or outer wall is heated. The case of both walls being heated was not addressed, as no experimental data were available for this condition. Wu \textit{et al.} (2009)~\cite{wu2009experimental} and Mayer \textit{et al.} (2016)~\cite{mayer2016experimental} focused on the so-called critical heat flux (CHF), which aims to predict the location of maximum heat flux occurrence in narrow annuli.

The literature review shows that radius ratio, Prandtl number, and thermal boundary conditions (including heat flux ratio) have a strong effect on heat transfer in annular pipe flows. However, compared to plane channel and pipe flows, research on the turbulent thermal boundary layer in heated annular pipe flow and its influence on emerging scaling relations of the turbulent heat transfer has remained unexplored. The curved inner wall precludes the application of the conventional wall functions, particularly for wide-gap cases described by a small radius ratio. To address the challenge of high computational cost in high-resolution simulations of fully developed turbulent flow, a map-based stochastic approach is adopted, utilizing Kerstein's one-dimensional turbulence (ODT) model~\cite{kerstein1999,kerstein2001} as a standalone tool, extending preliminary work by Tsai et al.~\cite{tsai2022,tsai2023,tsai2025}. The present study is based on the model formulation for velocity boundary layers in turbulent annular pipes that has been validated in~\cite{tsai2026}. Both the stochastic model formulation and the theoretical analysis are now extended to passive temperature transport, focusing on a moderate and low Prandtl number fluid under various flow conditions and radius ratios. The bulk Reynolds number, $Re_{D_{\rm h}}=1.77\times 10^4$, is primarily employed for the validation of the ODT predictions, as this Reynolds number provides the most comprehensive reference data available for comparison. After establishing the model accuracy, the study is extended to significantly higher Reynolds numbers, up to $Re_{D_{\rm h}}=10^5$ and $10^6$. At these Reynolds numbers, the capability of ODT is demonstrated through predictions of the Nusselt number as well as first- and second-order thermal statistics, thereby extending the investigation well beyond the range currently accessible to DNS.

The rest of this paper is structured as follows. Section~\ref{sec:formulation} describes the formulation of the ODT model and the governing equations for heated annular pipe flow. Section~\ref{sec:results} gives the main results in terms of thermal properties, such as Nusselt number, and turbulent thermal statistics, with a focus on the effects of spanwise curvature and the Prandtl number. Finally, we conclude the paper in Section~\ref{sec:conclusions}.

\section{Model formulation}\label{sec:formulation}
\begin{figure}[!htb]
    \centering
    \includegraphics[width=1.0\textwidth]{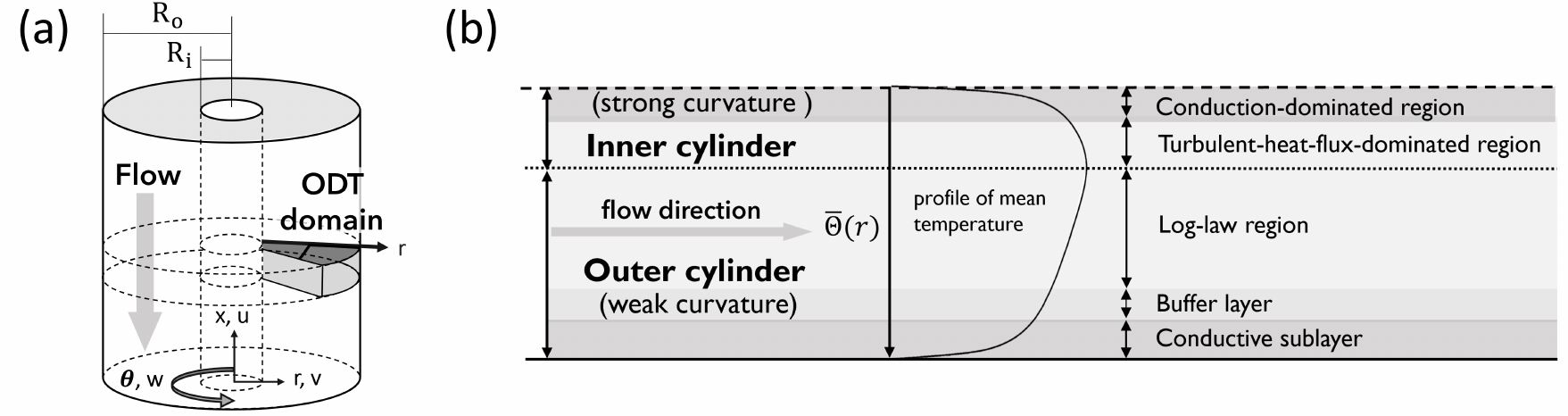}
    \caption{(a) Schematic drawing of the heated pressure-driven annular pipe flow configuration investigated. The radially oriented ODT domain represents an infinitesimal wedge. (b) Flow regions in the radial gap between the cylindrical inner and outer wall are shown with a schematic mean perturbation temperature profile $\overline{\Theta}(r)$. 
    }
    \label{fig:geometry}
\end{figure}

Figure~\ref{fig:geometry}(a) shows the geometry and coordinate system of the concentric annular pipe flow investigated. The coordinates $x$, $r$, and $\theta$ denote the axial, radial, and azimuthal directions, respectively, while $u$, $v$, and $w$ represent the velocity components in these directions. The computational domain, the ODT domain, is located in the radial direction, spanning the entire gap between the cylindrical inner and outer walls of the annular pipe. 

ODT~\cite{kerstein1999} is a dimensionally reduced flow model that aims to mimic the turbulence phenomena with a minimal representation of the interaction between simultaneously evolving momentum and scalar property profiles. The model distinguishes molecular diffusive transport from turbulent advective transport processes and aims to resolve all relevant scales of a turbulent flow along a single physical coordinate. Turbulent advection is modeled by a stochastic process, which is realized in practice as a stochastically sampled sequence of \textit{eddy events}. The \textit{eddy events} are formulated by a physical mapping operation utilizing the triplet map~\cite{kerstein1999,kerstein2001} as a building block model for map-induced turbulent fluxes in the radial direction. Eddy event locations, sizes, and occurrence rates are dynamically and locally determined based on the evolving momentum field of the flow. Due to its one-dimensional nature, ODT offers a significant reduction of computational cost compared to DNS (e.g.~\cite{klein2022stab}) for similar radial resolution and dynamical fidelity. Although ODT cannot replace DNS, it is complementary to large-eddy simulation (LES) and Reynolds-averaged Navier--Stokes simulation (RANS) as ODT does not require closure and closure modeling assumptions. ODT offers predictive capabilities for heat and momentum transfer as collective statistical results of resolved and evolving fine-scale flow features. 

\begin{figure}[!htb]
    \centering
    \includegraphics[width=1.0\textwidth]{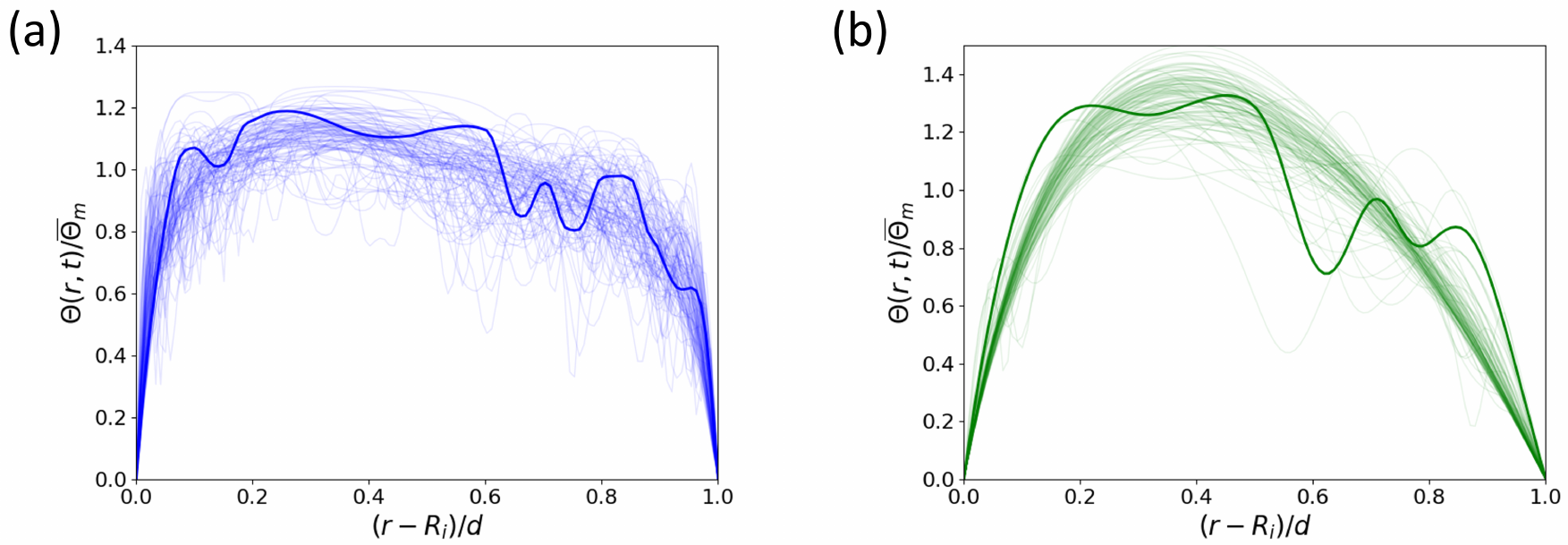}
    \caption{Instantaneous temperature profiles at different simulation times for Prandtl number (a) $Pr=0.71$ and (b) $Pr=0.025$, bulk Reynolds number $Re_{D_{\rm h}}=1{,}77\times 10^4$ and radius ratios $\eta=0.1$.
    }
    \label{fig:instant}
\end{figure}

Heated annular pipes are modeled analogously to~\cite{tsai2026} by adopting the ODT model formulation for cylindrical geometry~\cite{lignell2018}. The flow is assumed to be incompressible and pressure-driven, with no-slip and thermal boundary conditions prescribed on the inner and outer cylindrical walls. Temperature is treated as a passive scalar, an assumption valid under low heating conditions where buoyant effects are negligible. Applying the notation from Klein et al. (2022)~\cite{klein2022}, the ODT governing equations for velocity and passive temperature evolution can be written as
\begin{align}
   \frac{\partial {u}_i}{\partial t} 
   +\sum_{t_{\rm e}}\mathcal{E}_{u}\,\delta\left(t-t_{\rm e}\right) &=
    \frac{1}{r}\frac{\partial}{\partial r}\bigg(r\nu\frac{\partial {u}_i}{\partial r}\bigg)
    -\frac{1}{\rho}\frac{{\rm d}P}{{\rm d}x}\,\delta_{ix}
    \; ,
    \label{eq:gov_vel}
  \\
   \frac{\partial \Theta}{\partial t} 
   + \sum_{t_{\rm e}}\mathcal{E}_{\Theta}\,\delta\left(t-t_{\rm e}\right) &=
    \frac{1}{r}\frac{\partial}{\partial r}\bigg(r\alpha\frac{\partial \Theta}{\partial r}\bigg)
    + u \,\frac{{\rm d}\overline{T}_{\rm w}}{{\rm d}x}
    \; . 
    \label{eq:gov_therm}
\end{align}
Here ${u}_i$, $i=x,r,\theta$, represents the model-resolved instantaneous velocity vector in cylindrical coordinates, where $u_x=u$, $u_r=v$, and $u_\theta=w$ correspond to the axial, radial, and azimuthal directions, respectively. $\delta_{ij}$ is the Kronecker delta, and ${\text{d}P}/{\text{d}x}$ denotes the prescribed mean pressure-gradient force density in the axial direction. Under the assumption of statistically stationary annular pipe flow with constant heat flux imposed at both the inner and outer cylindrical walls, the temperature field exhibits a linear increase in the axial direction $x$. Following \cite{klein2022}, the perturbation temperature difference is defined as $\Theta=\overline{T}_{\rm w}-T$, where $\overline{T}_{\rm w}$ is the mean wall temperature, which increase linearly with $x$, and $T$ is the physically measured temperature. ${{\rm d}\overline{T}_{\rm w}}/{{\rm d}x}$ is mean axial gradient of the wall temperature, which is constant only under specific canonical settings of the thermal boundary conditions~\cite{chung2003,bagheri2021,fukuda2020,tsai2022}. The fluid properties, mass density $\rho$, kinematic viscosity $\nu$, and thermal diffusivity $\alpha$, are assumed constant. Thermal diffusivity $\alpha$ is defined as $\alpha=\lambda/\left( \rho C_{\rm p}\right)$, where $\lambda$ is thermal conductivity and $C_{\rm p}$ is specific heat capacity. Time is denoted by $t$, while $t_{\rm e}$ indicates the stochastically sampled times at which eddy events occur. The Dirac delta function is represented by $\delta(t)$. The terms $\mathcal{E}_u$ and $\mathcal{E}_\Theta$ symbolically represent instantaneous flow profile modifications, which model the instantaneous microstructure modifications with the aid of turbulent advection events. Due to the model's one-dimensional formulation, an additional pressure fluctuation model is included~\cite{kerstein2001}, leading to a distinct formulation of $\mathcal{E}_u$ compared to $\mathcal{E}_\Theta$ for a selected turbulent eddy event leading to simultaneous modifications of velocity and temperature profiles. Whenever a stochastically sampled eddy event occurs, the radial profiles $u_i(r)$ and $\Theta(r)$ are modified over a selected spatial eddy length scale $l_{\rm e}$ around a randomly sampled radial eddy location $r_{\rm e}$ using a mapping procedure. Here, this mapping is given by the measure and section-length preserving triplet map TMB~\cite{lignell2018,klein2023}. Further details, including the sampling procedure, are skipped here but can be found in~\cite{kerstein1999}. For the mesh-adaptive implementation in cylindrical geometry, refer to~\cite{lignell2018}, and for the treatment of the passive scalar, see~\cite{klein2022}. The model calibration specific to annular pipe flow follows the model parameters setting in~\cite{tsai2022}, where the energy distribution parameter ${\alpha}_{\rm odt}=1/6$, the eddy rate parameter $C_{\rm odt}=5$, and the viscous suppression parameter $Z_{\rm odt}=400$ as previously calibrated in~\cite{tsai2022}. Figure~\ref{fig:instant} gives an example of the ODT simulation for instantaneous temperature profiles normalized by the mean perturbation temperature, $\Theta(r,t)/\overline{\Theta}_m$, for two Prandtl numbers, illustrating the variability of the temperature profile in the statistically stationary state due to temporal evolution under fully-developed, mildly turbulent flow conditions. 

For normalization purposes, the local friction velocity and local friction temperature at the inner and outer walls are defined as 
\begin{align}
    u_{\tau, \, \rm i/o} = \sqrt{ \frac{\tau_{\rm \, i/o}}{\rho} }
    \,  \quad\text{and}\quad
    \Theta_{\tau, \, \rm i/o} = \frac{\alpha}{u_{\tau, \, \rm i/o}}\left|\frac{\mathrm{d} \overline{\Theta}}{\mathrm{d} r}\right|_{{\rm wall}, \, \rm i/o} 
    \; .  
    \label{eq:frictionVelTemp}
\end{align}
Here, $u_{\tau, \, \rm o/i}$ denotes the friction velocity at the inner (subscript $\rm i$) and outer (subscript $\rm o$) walls, respectively. The wall shear stress, which defines the friction velocity, is expressed as $\tau_{\rm i/o} = \mu \left|{\mathrm{d}\overline{u}}/{\mathrm{d}r}\right|_{{\rm wall}, \, \rm i/o}$, where $\mu$ is the dynamic viscosity. $\Theta_{\tau, \, \rm o/i}$ represents the friction perturbation temperature at the respective wall. The normalized mean perturbation temperature is given by $\overline{\Theta}_{\rm o/i}^+ = \overline{\Theta} / \Theta_{\tau, \, \rm o/i}$, and the corresponding normalized radial coordinate is defined as $r^+ = \left(|r - R_{\rm o/i}| \,  u_{\tau, \, {\rm o/i}}\right) / \nu$.

The average wall heat flux, $q_{\rm w}$ is defined as 
\begin{equation}
    q_{\rm w}=\frac{q_{\rm o} + \eta q_{\rm i}}{1+\eta} 
    \; ,
    \label{eq:qw}
\end{equation}
where $q_{\rm i}$ and $q_{\rm o}$ are the wall heat fluxes at the cylindrical inner and outer annular pipe walls, respectively, and they are assumed to be constant in this study. The average friction velocity and average friction temperature are defined separately as 
\begin{equation}
    u_\tau=\sqrt{\frac{\tau_{\rm o} + \eta \tau_{\rm i}}{\rho(1+\eta)}}
    \,  \quad\text{and}\quad
    \Theta_\tau= \frac{q_{\rm w}}{\rho c_p u_\tau}  
    \; . 
    \label{eq:avegFriction}
\end{equation}
Here, $c_p$ is the specific heat at constant pressure. Average friction Reynolds number is then defined as $Re_{\tau}=u_{\tau}H/\nu$, where $H$ is the half-width of the radial gap of the annular pipe, $H=(R_{\rm o}-R_{\rm i})/2$. Accordingly, the normalized mean perturbation temperature is defined as $\overline{\Theta}_{\rm i/o}^{+}=\overline{\Theta}/\Theta_{\tau,\rm i/o}$, and the corresponding wall-normal coordinate is given by $r^+ = \left(|r - R_{\rm i/o}| \,  u_{\tau, \, {\rm i/o}}\right) / \nu$. 

\section{Results and discussion}\label{sec:results}
\subsection{Nusselt number}\label{sec:Nu}
Heat transfer in annular pipe flow is conventionally characterized using the Nusselt number ($Nu$), which represents the ratio of convective to purely conductive heat transfer across the wall. The annular configuration has two domain walls, which give rise to two Nusselt number definitions, a global and a local one. The bulk Nusselt number quantifies the overall heat transfer in an integral sense, whereas the local gradient-based Nusselt numbers are evaluated at the cylindrical inner and outer walls, respectively. These local values quantify the radial asymmetry in the wall heat fluxes that govern the thermal boundary layers and the temperature distribution across the radial gap.

\subsubsection{Bulk Nusselt number}\label{sec:bulkNu}
Following~\cite{fukuda2020}, the bulk Nusselt number is defined
\begin{equation}
    Nu \equiv \frac{q_{\rm w}}{\lambda\overline{\Theta}_{\rm m}/D_{\rm h}}=\frac{4Re_{\tau}Pr}{\overline{\Theta}_{\rm m}^+}
    \, \quad \text{with}\quad
    \overline{\Theta}_{\rm m} = \frac{\int\overline{u}(r)\,\overline{\Theta}(r)\,r\,{\rm d}r }{\int\overline{u}(r)\,r\,{\rm d}r }    
    \; .
    \label{eq:NuBulk}
\end{equation}
Here, $\overline{\Theta}_{\rm m}^+=\overline{\Theta}_{\rm m}/\Theta_\tau$ is the normalized mean perturbation temperature, where $\overline{\Theta}_{\rm m}$ is the mean perturbation temperature. 

\begin{figure}[!htb]
    \centering
    \includegraphics[width=0.8\textwidth]{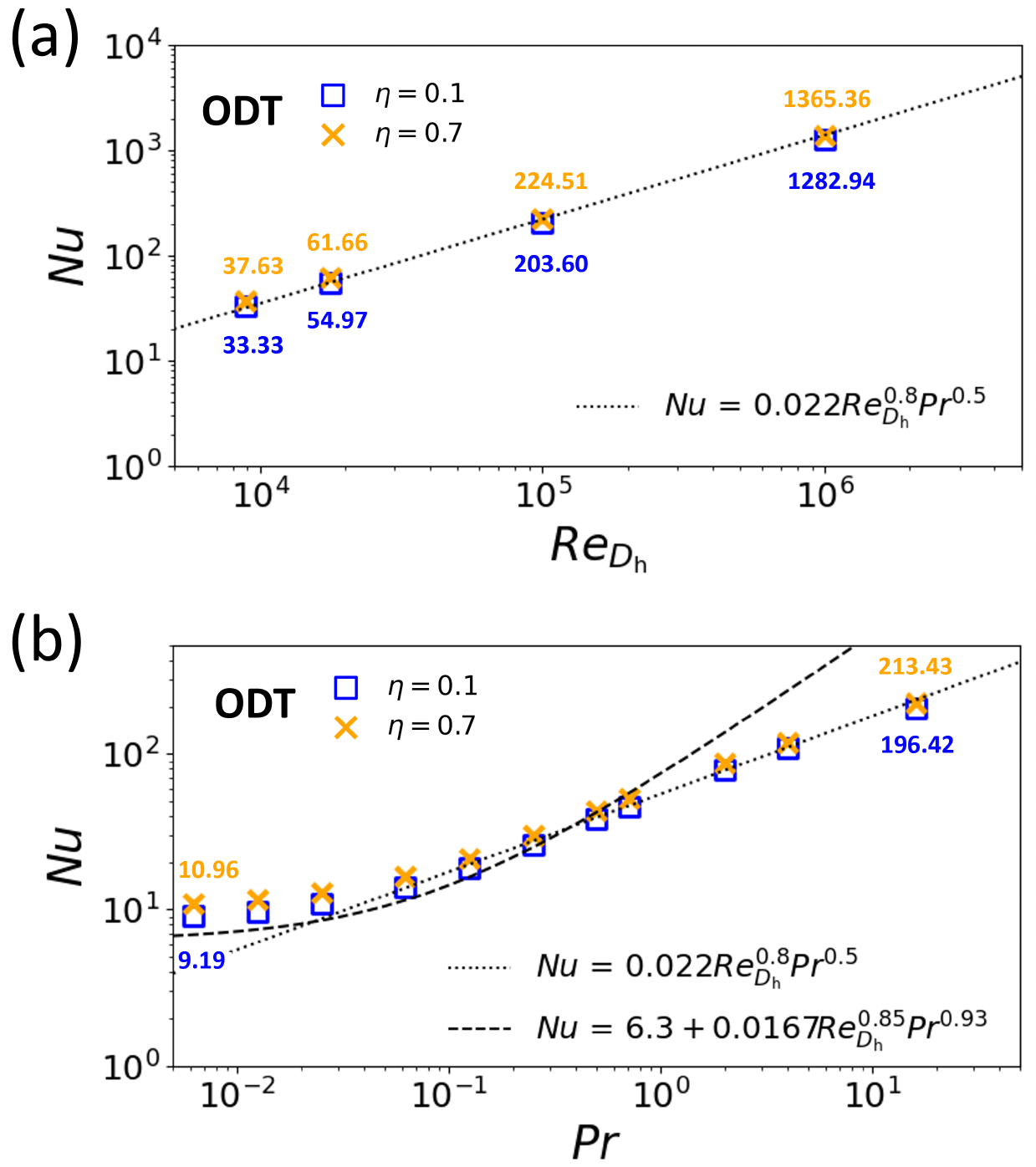}
    \caption{Bulk Nusselt number, $Nu$, as a function of (a) bulk Reynolds number, $Re_{D_{\rm h}}$, and (b) Prandtl number, $Pr$. Two radius ratios, $\eta=0.1$ and $0.7$, are investigated in (a) at fixed $Pr=1$ and in (b) at fixed $Re_{D_{\rm h}}=1{,}77\times 10^4$. The empirical correlation (dotted line)~\cite{kays1980} for Prandtl number close to unity and another empirical correlation (dashed line)~\cite{sleicher1975} for low-Prandtl-number are given for comparison.
    }
    \label{fig:BulkNu}
\end{figure}

The Nusselt number is compared against the empirical correlations~\cite{kays1980,sleicher1975}. The classical power-law correlation of~\cite{kays1980} (dotted line in Figure~\ref{fig:BulkNu}), namely
\begin{equation}
    Nu = 0.022 Re_{D_{\rm h}}^{0.8}Pr^{0.5}
    \; ,
    \label{eq:Nu_Kays}
\end{equation} 
which gives a reasonable prediction of the trend of the Prandtl number close to unity. The empirical correlation developed by~\cite{sleicher1975} (dashed line in Figure~\ref{fig:BulkNu}(b)), 
\begin{equation}
    Nu = 6.3 + 0.0167 Re_{D_{\rm h}}^{0.85}Pr^{0.93}
    \; ,
    \label{eq:Nu_Sleicher}
\end{equation}
is an approximation for the behavior of fluids at a very low Prandtl number.

Figure~\ref{fig:BulkNu}(a,b) presents the bulk Nusselt number as a function of bulk Reynolds number $Re_{D_{\rm h}}$ and the Prandtl number $Pr$, respectively. Bulk Reynolds numbers of $Re_{D_{\rm h}} = 8{,}900$, $1{,}77\times 10^4$, $10^5$, and $10^6$, together with Prandtl numbers in the range $Pr = 0.00625$–$16$, are considered for two radius ratios, $\eta = 0.1$ and $0.7$. The results are compared against established empirical correlations~\cite{kays1980,sleicher1975}. Overall, the bulk Nusselt number predicted by the ODT model follows the trend of the reference correlation~\cite{kays1980}, increasing monotonically with both Reynolds and Prandtl numbers. As shown in Figure~\ref{fig:BulkNu}(a), for $Re_{D_{\rm h}} \gtrsim 10^5$, the deviation between the ODT predictions and Kays' correlation remains small and is, in fact, within approximately $5\%$. Similarly, Figure~\ref{fig:BulkNu}(b) indicates that, for $Pr \geq 1$, the discrepancy is also small, generally below $10\%$, demonstrating good agreement in the fully developed turbulent flow regime. In contrast, larger deviations are observed for low Prandtl number cases, where $Pr < 1$, when compared with the applicable empirical correlation~\cite{sleicher1975}, with differences exceeding $20\%$. Furthermore, even at high Reynolds numbers, noticeable differences in $Nu$ can be discerned between the two radius-ratio cases ($\eta = 0.1$ and $0.7$) investigated. This indicates that curvature effects remain notable up to high asymptotic $Re_{D_{\rm h}} = 10^6$ for moderate to low $Pr$ and continue to influence heat transport in annular pipe flow.

\subsubsection{Local Nusselt number}\label{sec:localNu}
Following~\cite{bagheri2021}, the local gradient-based Nusselt number is defined based on the near-wall temperature gradient, given by
\begin{equation}
    Nu_{\rm i/o} = \frac{D_{\rm h}}{\overline{\Theta}_{\rm m}}\frac{{\rm d}\overline{\Theta} }{{\rm d} r}\bigg|_{\rm i/o} 
    \; .
    \label{eq:NuLocal}
\end{equation}
ODT facilitates the direct computation of the local Nusselt number thanks to the full-scale resolution of the radial property profiles, composed of convective- and diffusion-dominated regions.

\begin{figure}
    \centering
    \includegraphics[width=0.8\textwidth]{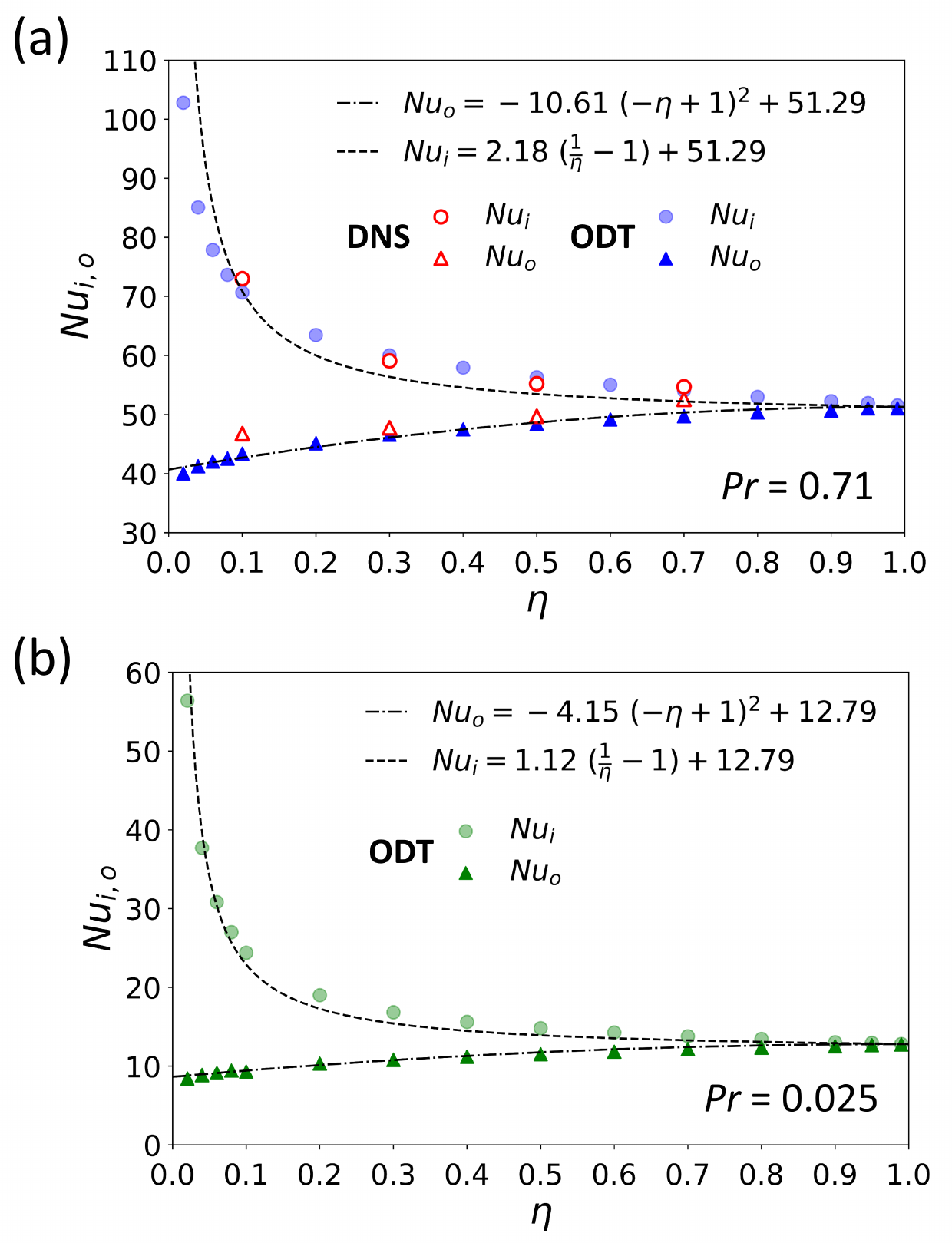}
    \caption{Local Nusselt number $Nu_{\rm i}$ and $Nu_{\rm o}$ at the cylindrical inner and outer wall, respectively, for Prandtl number (a) $Pr=0.71$ and (b) $Pr=0.025$, bulk Reynolds number $Re_{D_{\rm h}}=1{,}77\times 10^4$ and various radius ratios $\eta$. Reference DNS data~\cite{bagheri2021} is given for reference, but is available only for moderate $Pr=0.71$.
    }
    \label{fig:LocalNu}
\end{figure}

Figure~\ref{fig:LocalNu}(a,b) presents the local Nusselt numbers on the inner and outer walls, $Nu_{\rm i}$ and $Nu_{\rm o}$, respectively, as functions of the radius ratio $\eta$ for two different Prandtl numbers, $Pr=0.71$ and $Pr=0.025$, at a fixed Reynolds number of $Re_{D_{\rm h}}=1{,}77\times 10^4$. For moderate $Pr=0.71$, the ODT model captures the radius ratio dependence of the local heat transfer quite accurately when compared with reference DNS data~\cite{bagheri2021}, without the need for additional model parameter calibration. The maximum observed error between ODT and DNS is $3.12\%$ for the inner cylinder wall and $7.46\%$ for the outer wall, both occurring at the smallest radius ratio considered, $\eta=0.1$. These discrepancies are primarily attributed to finite curvature and finite Reynolds number effects, as discussed in~\cite{tsai2022,tsai2025}. Previous studies~\cite{klein2022} suggest that such modeling errors tend to diminish at higher Reynolds numbers, implying improved model performance in high Reynolds number regimes. For the lower Prandtl number case, $Pr=0.025$, no DNS reference data are currently available, so only a preliminary evaluation can be provided. The ODT-predicted trends of $Nu_{\rm i}$ and $Nu_{\rm o}$ resemble those seen in the $Pr=0.71$ case. For both $Pr$ investigated, the local Nusselt number is higher on the inner cylinder wall compared to the outer cylinder wall. This is attributed to the larger temperature gradient near the inner wall. The difference between the local Nusselt numbers diminishes as the radius ratio increases. This behavior is further analyzed in Figure~\ref{fig:meanSca} below for the mean temperature profile. 

Additionally, empirical fits to the ODT simulation results for both $Nu_{\rm i}$ and $Nu_{\rm o}$ are included in Figure~\ref{fig:LocalNu}(a,b), providing an analytical parameterization for the local heat transfer as a function of radius ratio $\eta$. The fitted expressions are given by
\begin{align}
    Nu_{\rm i} = m_{\rm i }\left( \frac{1}{\eta}-1 \right) + n
    \; ,
    \label{eq:Nu_i}
    \\
    Nu_{\rm o} = -m_{\rm o}\left( -\eta+1 \right)^2 + n
    \; ,
    \label{eq:Nu_o}
\end{align}
where $m_{\rm i}$, $m_{\rm o}$, and $n$ are empirical fitting coefficients. Their values are given in Figure~\ref{fig:LocalNu}.

\subsection{Mean temperature profile and thermal boundary layers}\label{sec:tempMean}
\subsubsection{Prandtl number effects}\label{sec:PrEffect}
\begin{figure}
    \centering
    \includegraphics[width=0.7\textwidth]{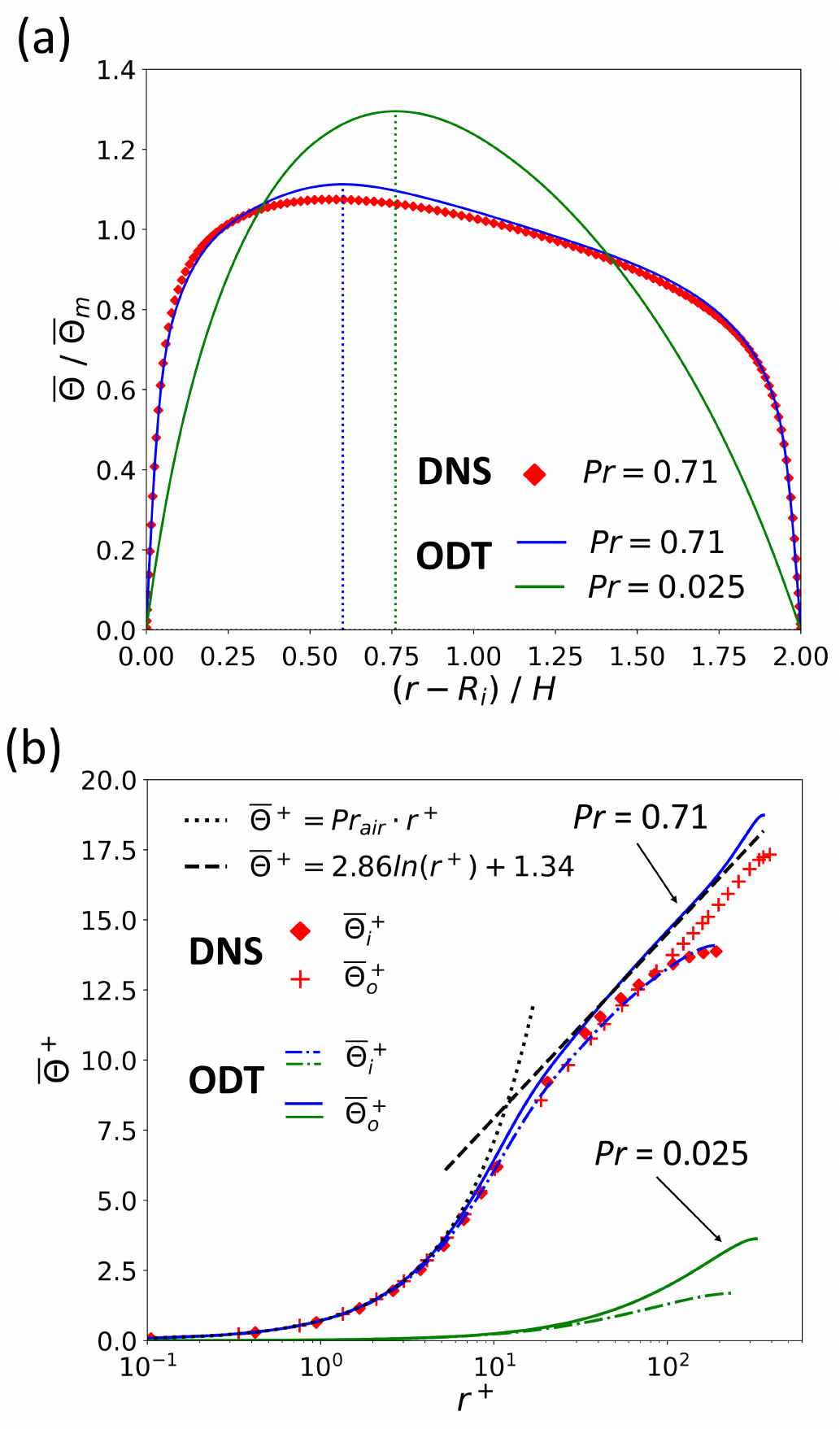}
    \caption{(a) Mean perturbation temperature profiles normalized by bulk mean temperature $\overline{\Theta}_{\rm m}$ and (b) boundary layer profiles of mean perturbation temperature over inner and outer cylinder walls for Prantl number $Pr=0.71$ and $Pr=0.025$ with fixed radius ratio $\eta=0.1$ and bulk Reynolds number $Re_{D_{\rm h}}=1{,}77\times 10^4$. The mean state is obtained as the long-time temporal (and ensemble) average of the instantaneous profiles shown in Fig. 2. Available reference DNS~\cite{bagheri2021} data is shown for comparison.
    }
    \label{fig:meanSca}
\end{figure}

Figure~\ref{fig:meanSca}(a) displays the mean perturbation temperature radial profiles predicted by ODT for two Prandtl numbers, $Pr=0.71$ and $Pr=0.025$, at moderate Reynolds number. For $Pr=0.71$, ODT predictions largely reproduce reference DNS~\cite{bagheri2021}, up to a $4\%$ overestimation of the magnitude of the maximum mean perturbation temperature, whereas the near-wall profile shapes and wall gradient appear to be captured even more accurately. Figure~\ref{fig:meanSca}(b) shows a law-of-the-wall plot over the inner and outer cylinder wall. The comparison of ODT with reference DNS and analytical wall functions demonstrates the model's applicability to forward modeling of turbulent boundary layers, providing further validation of the model formulation and its present calibration. For $Pr=0.025$, the temperature profile is close to the purely conductive solution for the relatively low $Re$ selected here. The reason is the large thermal diffusivity in relation to the kinematic viscosity, meaning heat is diffused more efficiently than momentum. As a result, thermal fluctuations induced by turbulence are more rapidly smoothed out, leading to a faster relaxation toward the conductive solution as can be inferred from Figure~\ref{fig:meanSca}(a). It would be illustrative to plot the purely conductive solution for comparison here since $\eta = 0.1$ is fixed. Moreover, the influence of wall curvature is clearly evident in both Prandtl number cases, regardless of whether conduction or turbulent advection dominates. The asymmetry in the temperature distribution across the radial gap highlights how spanwise curvature affects thermal transport. Notably, the peak of the mean perturbation temperature is located closer to the cylindrical inner wall, which is characterized by convex curvature and a smaller curvature radius than the outer wall. This radial asymmetry of the mean temperature profile results in a thinner thermal boundary layer adjacent to the inner wall and a correspondingly thicker boundary layer near the outer wall in line with the radial asymmetry in the local Nusselt numbers as discussed above (Figure~\ref{fig:LocalNu}).

Figure~\ref{fig:meanSca}(b) shows the thermal boundary layer at the inner and outer cylinder wall, respectively, for both Prandtl numbers and normalized to friction units, compensating for the wall gradient up to a predefined factor of $Pr$. It is found that the thermal boundary layer can be reasonably well described using a linear conductive sublayer and logarithmic regions, consistent with classical boundary layer theory. However, this simplified approach does not fully capture the behavior on the inner wall, where deviations from the assumed planar-like profile are more apparent. This discrepancy, particularly the influence of wall curvature and its interplay with radial transport, is discussed next.

\subsubsection{Radius ratio effects}\label{sec:ReEffect}
\begin{figure}
    \centering
    \includegraphics[width=0.7\textwidth]{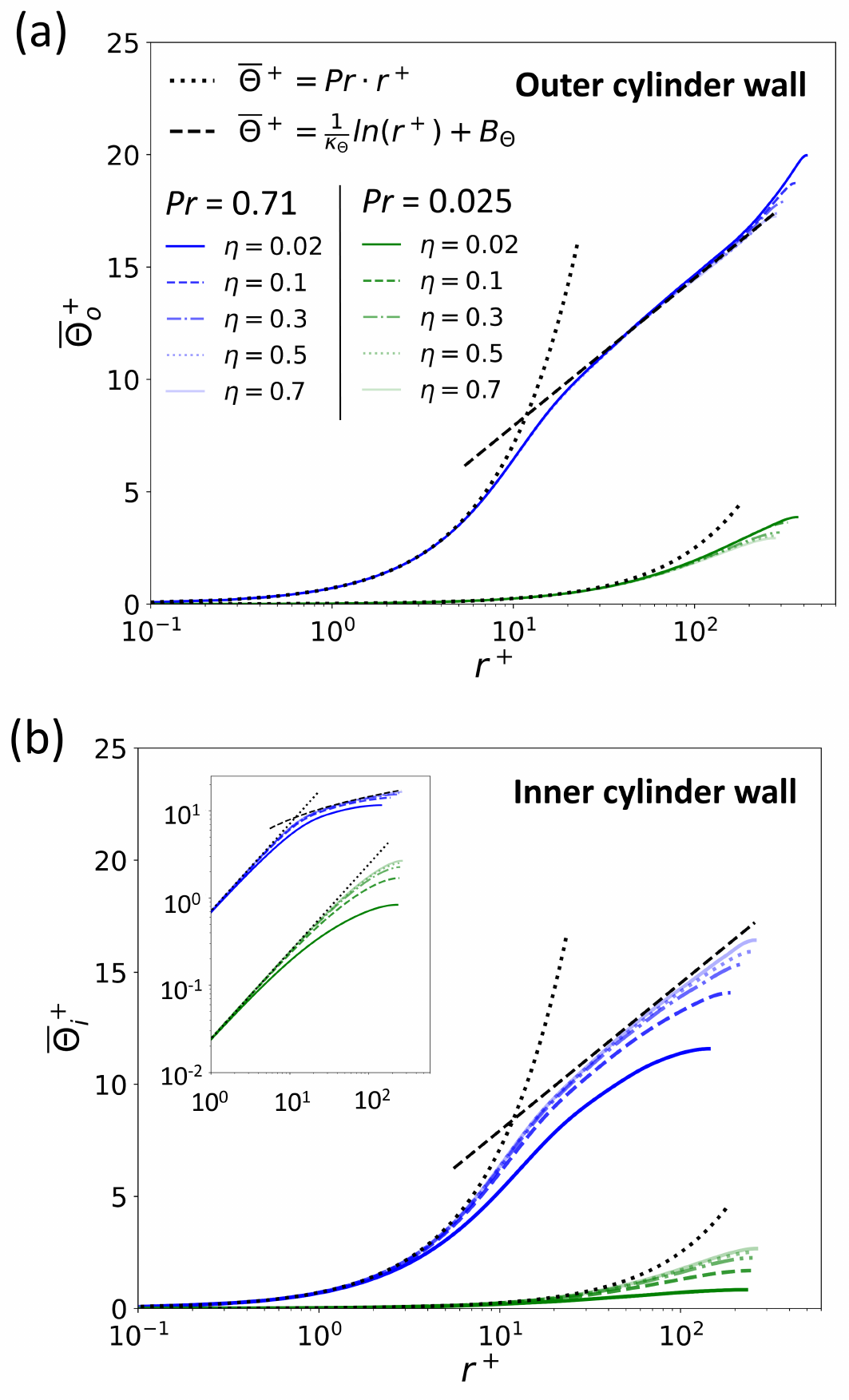}
    \caption{Thermal boundary layer over the (a) outer and (b) inner cylinder wall for various radius ratios, $\eta=0.02$ to $0.7$, and two Prandtl numbers, $Pr=0.71$ and $0.025$, with fixed bulk Reynolds number $Re_{D_{\rm h}}=1{,}77\times 10^4$.
    }
    \label{fig:MeanTempBL}
\end{figure}

Figures~\ref{fig:MeanTempBL}(a,b) extend the analysis to the thermal boundary layer across a range of radius ratios, from $\eta=0.02$ to $\eta=0.7$. In Figure~\ref{fig:MeanTempBL}(a), it is observed that wall curvature is much less influential at the outer cylinder wall since all normalized profiles collapse on the same master profile given by the conventional, quasi-planar law-of-the-wall. For $Pr=0.71$, the conductive sublayer exhibits a linear profile and extends up to $r^+\approx5$. By contrast, for $Pr=0.025$, a longer conductive sublayer is observed, extending up to $r^+=30$. In this region, molecular heat transfer dominates, similar to the role of viscous effects in the momentum boundary layer. For the $Pr=0.71$ case, the logarithmic region of the thermal boundary layer begins around $r^+\approx30$ and follows a logarithmic law, $\overline{\Theta}^+ = \frac{1}{\kappa_{\Theta}} \ln \left( r^+ \right) + B_{\Theta}^+$, characterized by the thermal von~Kármán constant $\kappa_\Theta=0.35$ and an additive constant $B_\Theta=1.34$~\cite{bagheri2021}, shown by the dashed line in Figure~\ref{fig:MeanTempBL}. For $Pr=0.025$, no logarithmic region can be discerned since the Reynolds number of $Re_{D_{\rm h}}=1{,}77\times 10^4$ is too low to push the mean profile notably away from the conductive solution.

Figure~\ref{fig:MeanTempBL}(b) shows that the wall curvature effect becomes pronounced at the inner cylinder wall, leading to significant variations in the boundary layer profiles for both Prandtl numbers. Under these conditions, the conventional logarithmic region is no longer clearly identifiable and fails to adequately represent the thermal boundary layer for finite values of $\eta$, $Pr$, and $Re_{D_{\rm h}}$, despite a visible tendency for emergence of the planar law-of-the-wall as $\eta\to 1$ for moderate $Pr=0.71$. 

\subsection{Boundary layer analysis}\label{sec:tempBL}
To quantify the effects of radius ratio, Prandtl and Reynolds number variation, the thermal boundary layer is further analyzed by extending the theoretical analysis presented in~\cite{tsai2026} to heat transport. The key idea is to divide the boundary layer over the inner cylindrical wall into two distinct regions: a conduction-dominated region and a turbulent-heat-flux-dominated region, as illustrated in Figure~\ref{fig:geometry}(b). 

\subsubsection{Conduction-dominated region}\label{sec:conductiveRegion}
Assuming a fully developed flow, all statistical properties remain constant in both the axial and azimuthal directions. As a result, only the steady-state balance equations governing the radial variation of the flow's statistical moments need to be considered. By concentrating on the mean, such as the first-order statistical moment, ODT can be effectively employed as a stand-alone model to provide flow statistics for an extended range of parameter values by Reynolds-averaging of the ODT solution, by temporal and ensemble averaging of autonomous flow realizations cycling through different random number sequences in the sampling procedure. The converged statistical moments solve the Reynolds-averaged ODT equations that are consistent with the Reynolds-averaged form of the full-order primitive equations. This leading-order statistical consistency between the Reynolds-averaged ODT and Navier-Stokes equations, in turn, enables the evaluation of the radial balance equations for axial velocity and perturbation temperature, as shown in~\cite{tsai2023,tsai2025}. Applying the Reynolds-averaging procedure to Equation~\eqref{eq:gov_therm}, followed by rearrangement and a single integration over the radial coordinate $r$, yields the heat flux balance equation, derived analogously to the shear stress balance equation as
\begin{align}
    \overline{\Theta^\prime v^\prime} - \alpha\,\frac{{\rm d} \overline{\Theta}}{{\rm d} r} = r\,\overline{u}\,\frac{{\rm d}\overline{T}_{\rm w}}{{\rm d}x} + \frac{C_1}{r}
    \; \quad\text{with}\quad
    C_1=\,\frac{1}{\rho\, C_p}\,\frac{q_{\rm i}\, R_{\rm o} - q_{\rm o}\, R_{\rm i}}{R_{\rm i}/R_{\rm o} - R_{\rm o}/R_{\rm i}} 
    \; . 
    \label{eq:gov_balance}
\end{align}
Here, $C_1$ is an integration constant that can be expressed in terms of the heat fluxes at the inner and outer cylindrical walls. $C_p$ is the heat capacity.

In the region very close to the wall, the conductive term, $\alpha\,\left({{\rm d}\overline{\Theta}}/{{\rm d} r}\right)$, significantly outweighs the turbulent transport term, $\overline{\Theta^\prime v^\prime}$. Assuming conduction dominates, meaning that $\overline{\Theta^\prime v^\prime} \ll \alpha\,\left({{\rm d} \overline{\Theta}}/{{\rm d} r}\right)$, and considering small but finite radius ratios ($\eta\ll1$), the normalized form of Equation~\eqref{eq:gov_balance} in friction units can be simplified to the following analytical expression,
\begin{align}
    \overline{\Theta}^+(r^+) = Pr \, R_{\rm i}^+\,\ln\left( \frac{r^+ + R_{\rm i}^+}{R_{\rm i}^+} \right)
    \; \quad\text{with}\quad
    R_{\rm i}^+=\frac{R_{\rm i}\,u_{\tau,\rm i}}{\nu} 
    \; .  
    \label{eq:cond_dom}
\end{align}
Here, $R_{\rm i}^+$ is not a constant, but rather a parameter that depends on the wall geometry and the state of the turbulent flow, meaning that it depends on the curvature radius and the wall-shear stress at the cylindrical inner wall. It is observed that the region near the wall does not follow a purely linear behavior. Taking the Taylor expansion of $\ln(\left(r^++R_i^+\right)/R_i^+)$ to lowest order around $r^+/R_i^+=0$ yields $\ln((r^++R_i^+)/R_i^+)\approx r^+/R_i^+$ such that the conventional linear conductive sublayer, $\overline{\Theta}^+(r^+)=Pr\, r^+$, is recovered in the vicinity of the cylindrical inner wall when the curvature radius is not much larger than the conductive and viscous length scales. 

\subsubsection{Turbulent-heat-flux-dominated region}\label{sec:turbHeatFluxRegion}
The same approach described above is applied to the turbulent-heat-flux-dominated region, with the difference that the total heat flux is now assumed to be predominantly transported by turbulent heat flux, such that $\alpha\,\left({{\rm d} \overline{\Theta}}/{{\rm d} r}\right) \ll \overline{\Theta^\prime v^\prime}$. For small radius ratios ($\eta\ll 1$), the integration constant $C_1$ can be simplified by considering the large-gap limit, $R_{\rm o} \gg R_{\rm i}$. For the rest, the radial dependence is kept in place. To model the turbulent transport term, a gradient-flux approximation is employed,
\begin{equation}
    \overline{\Theta^\prime v^\prime} \simeq -\alpha_t \frac{\rm d \overline{\Theta}}{{\rm d} r}
    \; .
    \label{eq:turbFluxApprox}
\end{equation} 
The turbulent thermal diffusivity is therefore defined as 
\begin{align}
    \alpha_t = -\frac{\overline{\Theta^\prime v^\prime}}{{\rm d} \overline{\Theta}/{\rm d} r}
    \; ,
    \label{eq:turbFlux}
\end{align} 
in analogy to the turbulent eddy viscosity~\cite{boersma2011},
\begin{equation}
    \nu_t = -\frac{\overline{u^\prime v^\prime}}{{\rm d} \overline{u}/{\rm d} r}
    \; .
    \label{eq:turbEddyViscosity}
\end{equation} 
Both $\alpha_t$ and $\nu_t$ contribute to the definition of the turbulent Prandtl number, 
\begin{equation}
    Pr_t=\frac{\nu_t}{\alpha_t}
    \; 
    \label{eq:turbPr}
\end{equation}
The eddy viscosity $\nu_t$ can be parameterized using mixing-length theory, while $\alpha_t$ can subsequently be determined using the Reynolds analogy, provided $Pr_t$ is known based on available data.

An expression for the turbulent eddy viscosity was previously proposed by~\cite{boersma2011}, taking the form $\nu_t=\epsilon_1 \sqrt{{\tau_{\rm i}}/{\rho}}\, (r-R_{\rm i})$, where $\epsilon_1$ is an unknown proportionality constant that must be determined from data. This mixing-length formulation implies the presence of nonlocal curvature effects, influencing both the conductive sublayer and the logarithmic region of the thermal boundary layer near the cylindrical inner wall. Most importantly, it is suggested that turbulent eddy sizes do not scale linearly with wall distance, but instead scale with the square root of the distance from the curved surface. This scaling can be both physically and mathematically motivated by enforcing area (measure) conservation during radial displacement of fluid parcels in an incompressible, constant-property flow. Although these arguments were not explicitly elaborated in~\cite{boersma2011}, the mathematical implications for model well-posedness were discussed. The mixing-length model by~\cite{boersma2011} remains consistent with the statistical average momentum transport, as confirmed by their DNS. It is worth noting that, the assumptions embedded in the reference mixing length model~\cite{boersma2011} align with the mass conservation and the measure-preserving properties encoded in the formulation of instantaneous radial mappings in ODT~\cite{lignell2018}, which underpin the construction of $\mathcal{E}_u$ and $\mathcal{E}_\Theta$ in Equations~\eqref{eq:gov_vel} and~\eqref{eq:gov_therm}, respectively. Since statistical moments are the collective effect of physical mappings across a range of scales, it follows from the building-block-based construction of ODT that the stochastic model must also reproduce the low-order statistical properties of the mixing-length model.

Substituting the mixing-length expression into Equation~\eqref{eq:gov_balance}, followed by normalization using friction units, yields the following analytical expression for the mean perturbation temperature profile in the convection-dominated (turbulent-heat-flux-dominated) region, as
\begin{equation}
    \overline{\Theta}^+(r^+) = \frac{Pr_t}{\epsilon_1^+}\,\ln\left( \frac{r^+}{r^+ + R_{\rm i}^+} \right) + D_1^+ 
    \; .
    \label{eq:turbHeatFlux_dom}
\end{equation}
Here, $\epsilon_1^+$ and $D_1^+$ are parametrization coefficients that can be determined from the ODT simulation data. The resulting mean perturbation temperature profile follows a logarithmic form with a shifted argument that explicitly depends on the Reynolds number. This dependence implies that a universal master profile does not exist for walls with significant spanwise curvature, which marks a clear departure from the classical law-of-the-wall on planar or weakly curved surfaces.

\subsubsection{Estimation of the turbulent Prandtl number}\label{sec:turbPr}
\begin{figure}
    \centering
    \vspace{-60pt}
    \includegraphics[width=0.7\textwidth]{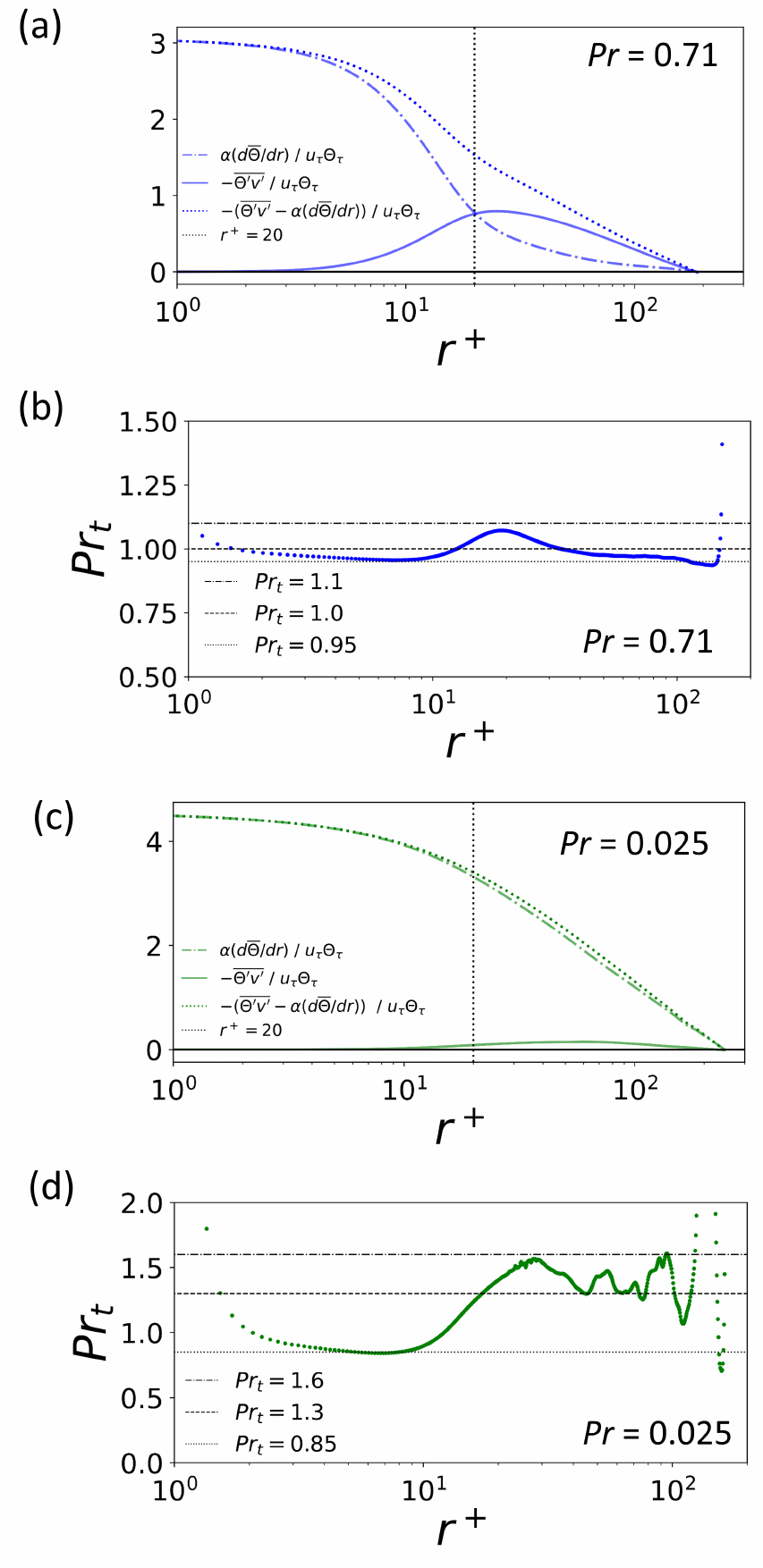}
    \caption{Normalized heat fluxes profiles, (a,c), and corresponding radial profile of the turbulent Prandtl number, (b,d), across the boundary layer over the cylindrical inner wall for Prandtl numbers (a, b) $Pr=0.71$ and (c, d) $0.025$ for fixed radius ratio $\eta=0.1$ and bulk Reynolds number $Re_{D_{\rm h}}=1{,}77\times 10^4$ as simulated with ODT.
    }
    \label{fig:TurbPr}
\end{figure}

Figure~\ref{fig:TurbPr}(a,c) shows the radial profiles of normalized conductive heat flux $q_{\rm{c}} = -\alpha\,\left({{\rm d}\overline{\Theta}}/{{\rm d} r}\right)/ u_{\tau} \Theta_{\tau}$, the normalized turbulent heat flux $q_{\rm{t}} = \overline{\Theta^\prime v^\prime} / u_{\tau} \Theta_{\tau}$, which is directly computed in ODT by averaging the collective map-induced transport per unit time interval (see~\cite{klein2022}, Appendix C), and the total heat flux, calculated as $q_{\rm{tot}} = q_{\rm{c}} + q_{\rm{t}}$. These profiles are shown for Prandtl numbers $Pr=0.71$ and $0.025$, with fixed radius ratio $\eta=0.1$, and a bulk Reynolds number $Re_{D_{\rm h}}=1{,}77\times 10^4$. For $Pr=0.71$ case, there is a clear separation between the regions dominated by conductive and turbulent thermal transport. Specifically, the conductive heat flux governs the temperature profile in the near-wall region, but for $r^+\geq5$, the conductive dominance decreases rapidly in line with established turbulent boundary layer properties. By contrast, the turbulent heat flux becomes the primary contributor to heat transfer for $r^+>20$, consistent with the development of enhanced mixing due to turbulent eddies at the onset of the turbulent logarithmic layer. For $Pr=0.025$, the conductive heat flux remains dominant across almost the entire wall-normal range, owing to the significantly higher thermal diffusivity of the fluid. Even beyond $r^+>20$, where turbulent transport begins to appear, its contribution to the overall heat flux remains comparatively weak.

The radial profile of the turbulent Prandtl number over the cylindrical inner wall, computed from the ODT-resolved Reynolds shear stress and turbulent heat flux, is shown in Figures~\ref{fig:TurbPr}(b,d). For $Pr=0.71$, an approximately constant value of $Pr_t\simeq0.95$ describes the near-unity values reasonably well across almost the entire boundary layer. For $Pr=0.025$, a larger variability is discerned, leading to larger uncertainty of the $Pr_t$ value to be selected. Slightly larger than unity values are robustly obtained off the wall, with $Pr_t\simeq1.3$ being a reasonable approximation that is applicable throughout the logarithmic region ($r^+>30$).

\subsubsection{Reynolds and Prandtl number effects}\label{sec:RePrEffect}
Figure~\ref{fig:BLcoeff}(a) shows the fully-developed turbulent thermal boundary layer at the cylindrical inner wall for Reynolds numbers ranging from $Re_{D_{\rm h}}= 1{,}77\times 10^4, 10^5, 10^6$. ODT simulation results are compared to fitted analytical expressions for the conduction and turbulent-heat-flux dominated regions for Prandtl numbers $Pr=0.71$ and $0.025$, keeping the radius ratio fixed at $\eta=0.1$. Analytical profiles are given by Equations~\eqref{eq:cond_dom} and \eqref{eq:turbHeatFlux_dom}, already utilizing the applicable turbulent Prandtl number $Pr_t$ obtained from the ODT data as shown in Figures~\ref{fig:TurbPr}(b,d).

In the conductive-dominated region, Equation~\eqref{eq:cond_dom} provides an excellent representation of the near-wall temperature profile. This is evidenced by the collapse of all simulation data onto the dashed master curve for various $Re$ per selected $Pr$. Likewise, Equation~\eqref{eq:turbHeatFlux_dom} fairly accurately describes the turbulent logarithmic region of the thermal boundary layer over the cylindrical inner wall for all $Re$ and $Pr$ investigated. The applicability extends to the outer layer within the standalone application of ODT, consistent with the bulk profile shown in Figure~\ref{fig:meanSca}(a). Interestingly, $Pr=0.025$, even though the region $r^+\ge20$ remains influenced by heat conduction, Equation~\eqref{eq:turbHeatFlux_dom} still parameterizes the radial profile shape of the perturbation temperature. Similar to the development of the velocity boundary layer in the momentum field, increasing the Reynolds number leads to a thicker thermal boundary layer in both Prandtl number cases. Notably, in the low Prandtl number case ($Pr=0.025$), a distinct logarithmic region in the temperature profile becomes increasingly evident at high asymptotic Reynolds numbers ($Re_{D_{\rm h}}>10^5$), as also reported in~\cite{ould2010}. This behavior highlights the growing influence of turbulent mixing in the thermal transport process, even in regimes where conductive heat transfer is still significant near the wall.

\subsubsection{Parameterization of radius ratio effects}\label{sec:paraRadius}
\begin{figure}
    \centering
    \vspace{-70pt}
    \includegraphics[width=0.7\textwidth]{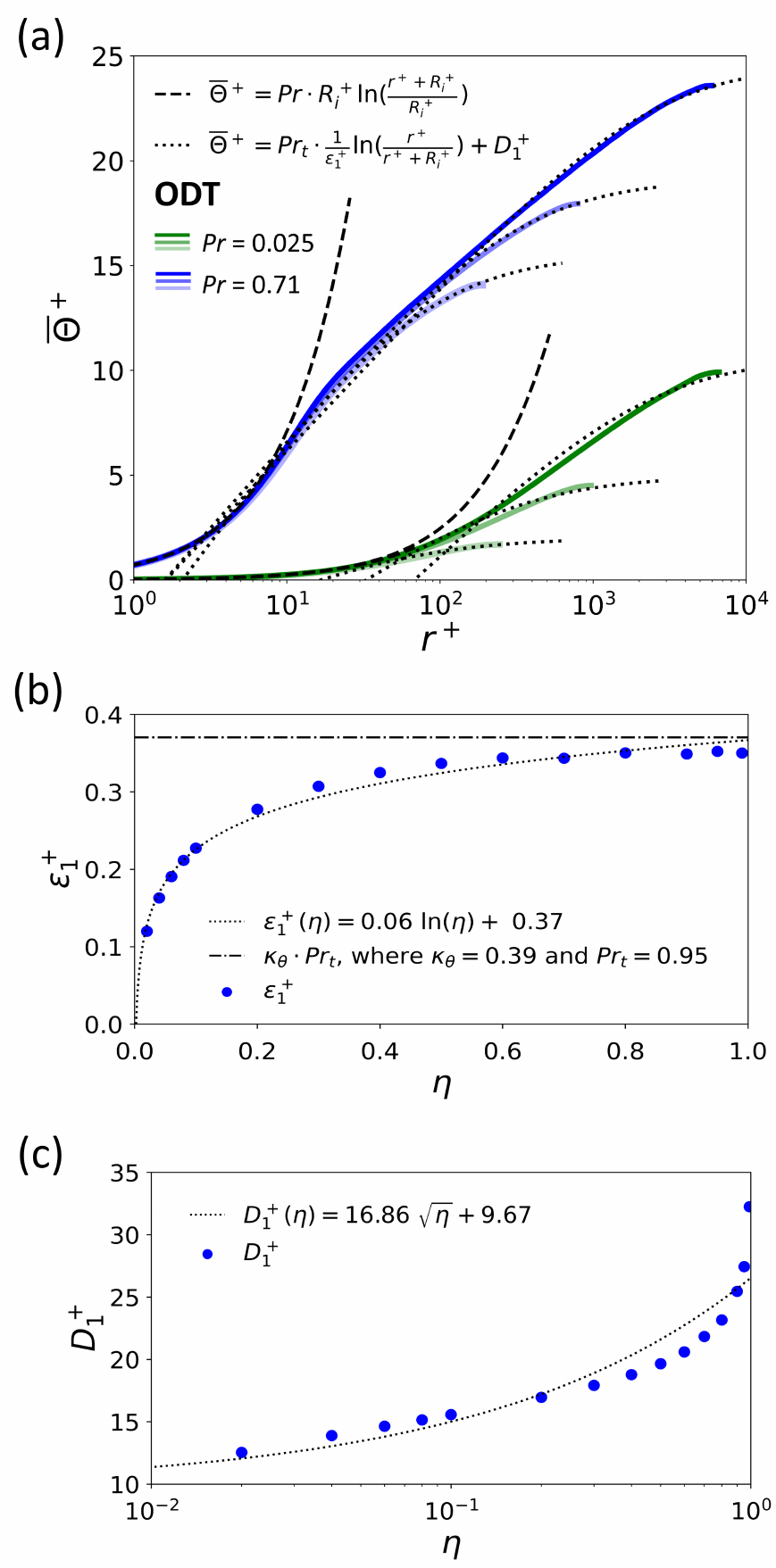}
    \caption{(a) Boundary layer profiles of the mean perturbation temperature over the cylindrical inner wall for various bulk Reynolds numbers $Re_{D_{\rm h}}= 1.77\times 10^4, 10^5$ and $10^6$ (shading of lines from light to dark) investigating two Prandtl numbers. $Pr=0.71$ (blue) and $0.025$ (green), respectively, for the fixed radius ratio $\eta=0.1$. Radius ratio dependence of the profile parameter (b) $\epsilon_1^+$ and (c) $D_1^+$ obtained by fitting the analytical expressions to the ODT simulation results in the vicinity of the cylindrical inner wall.  
    }
    \label{fig:BLcoeff}
\end{figure}

The radial profile shapes are uncertain only with respect to radius ratio effects that are discussed next, as these are governing radial asymmetry and integration constants. Figures~\ref{fig:BLcoeff}(b,c) show the variation of the constant coefficients $\epsilon^+_1$ and $D^+_1$ across a wide range of radius ratios $\eta$. As the radius ratio increases, wall curvature effects diminish and the coefficient $\epsilon^+_1$ decreases monotonically, approaching an asymptotic value that closely aligns with the reference von K\'arm\'an constant $\kappa_{\theta}$ for channel flow, as reported in~\cite{klein2022}. By contrast, the integration constant $D^+_1$ increases with $\eta$. This behavior highlights the significant influence of geometric curvature on the offset term in the temperature profile. The two constant coefficients are empirically expressed as functions of $\eta$ and given by
\begin{align}
    \epsilon^+_1 ( \eta ) = 0.06 \ln( \eta ) + 0.37
    \; ,
    \label{eq:epsilon_1}
    \\
    D^+_1 ( \eta ) = 16.86 \sqrt{\eta} + 9.67
    \; .
    \label{eq:D1}
\end{align}

\subsection{Turbulent heat flux}
\begin{figure}
    \centering
    \includegraphics[width=0.7\textwidth]{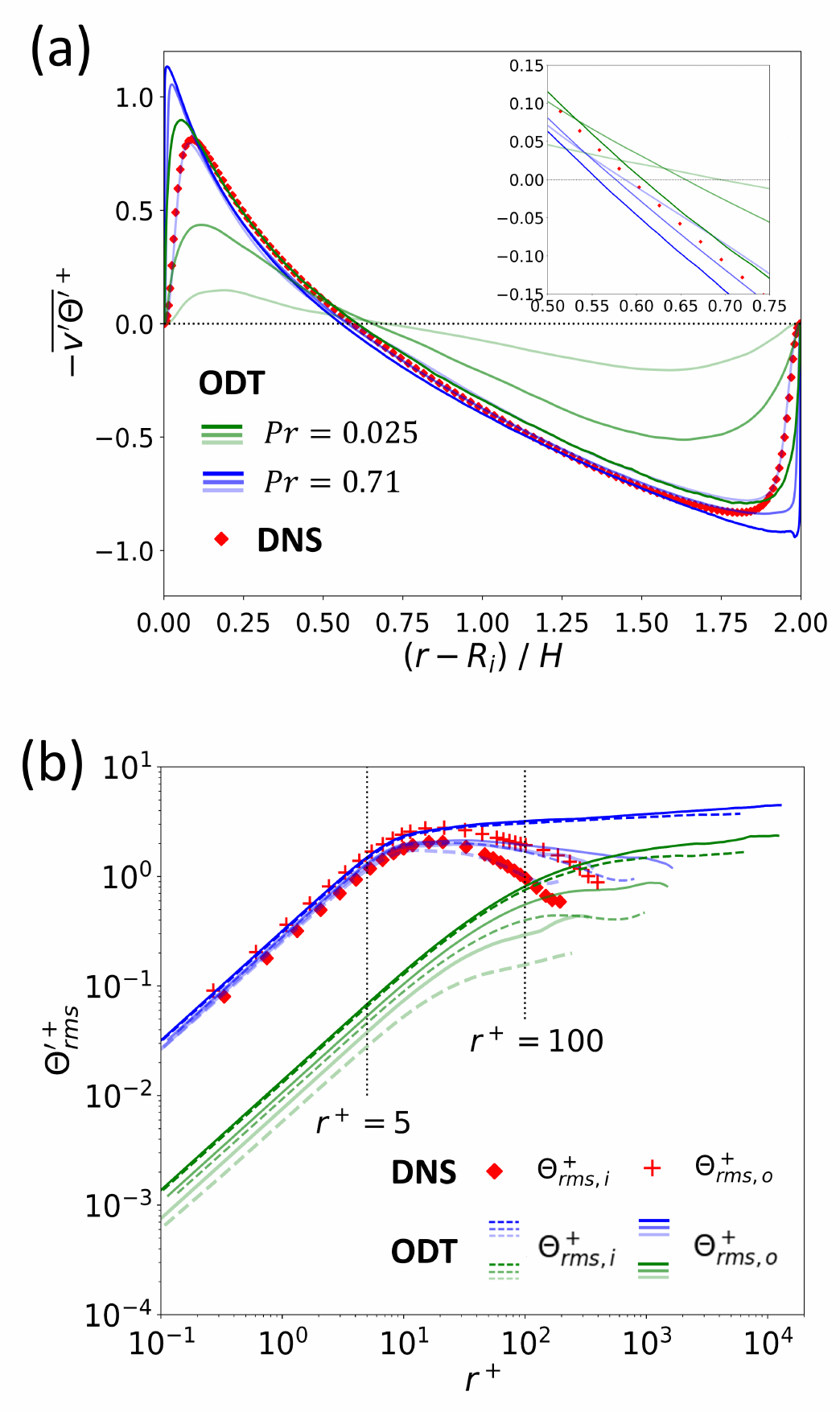}
    \caption{Radial profiles of the (a) normalized turbulent heat flux in radial direction $-\overline{v^\prime \Theta^\prime}^+$ and (b) standard deviation (RMS) of the normalized temperature fluctuations $\Theta^+_{\rm rms}$ for various bulk Reynolds numbers $Re_{D_h}=1.77\times 10^4, 10^5$ and $10^6$ (shading of lines from light to dark) investigating two Prandtl numbers, $Pr=0.71$ (blue) and $0.025$ (green), respectively, for the fixed radius ratio $\eta=0.1$.
    Reference DNS data from~\cite{bagheri2021} for $Re_{D_h}=1.77\times 10^4$ and $Pr=0.71$ using the same radius ratio are given for comparison.
    }
    \label{fig:TurbFlux}
\end{figure}

To further understand the influence of the curvature on heat transport, the radial turbulent heat flux is examined in the following. The mean temperature profiles indicate the overall thermal distribution, whereas the turbulent heat flux quantifies the turbulent transport of heat across the annular gap. Figure~\ref{fig:TurbFlux}(a) presents the radial turbulent heat flux, denoted as $-\overline{v^\prime\Theta^\prime}^+$, for various Reynolds and Prandtl numbers at a fixed radius ratio of $\eta=0.1$. The result for the case with $Re_{D_{\rm h}}=1{,}77\times 10^4$ and $Pr=0.71$ demonstrates good agreement with the reference DNS data~\cite{bagheri2021}, providing additional validation for ODT and demonstrating that the map-induced radial flux correlates with the capturing of the turbulent boundary layer radial profile. Radial asymmetry is observed for the turbulent heat flux that resembles the behavior of the Reynolds shear stress~\cite{bagheri2020}. For $Pr=0.71$, the normalized turbulent heat flux has a larger magnitude than for $Pr=0.025$. Furthermore, an increase in Reynolds number leads to a noticeable amplification in the turbulent heat flux, indicating enhanced convective transport. It is observed that the Reynolds number has a relatively minor effect on the location of the zero-crossing for $Pr=0.71$. However, for $Pr=0.025$, as the Reynolds number increases, the zero-crossing shifts toward the cylindrical inner wall.

\subsection{RMS temperature}
Figure~\ref{fig:TurbFlux}(b) presents the normalized root-mean-square (RMS) of temperature fluctuation, denoted as $\Theta^+_{\rm rms}$, highlighting its sensitivity to both the Reynolds number and the Prandtl number. The results indicate that temperature fluctuations are generally more significant near the outer wall than near the inner wall. Moreover, the disparity between the temperature variations at the inner and outer walls becomes more pronounced in regions farther from the walls. This asymmetry is particularly significant at lower Reynolds and Prandtl numbers. Compared to available reference DNS~\cite{bagheri2021}, ODT underestimates the peak value of the temperature fluctuation, which is observed at $r^+\approx15$, but captures the main statistical features. For higher Reynolds number, the temperature RMS radial profile develops a broad plateau toward the bulk, diminishing the near-wall peak discernible for $Pr=0.71$. For $Pr=0.025$, the RMS increases monotonically with wall distance for all $Re$ investigated, but tends to develop a plateau in the bulk for high asymptotic Reynolds numbers. 

Notice that for $Pr=0.025$, no reference DNS is available for comparison. Specifically, the model predicts overall less normalized temperature fluctuations relative to $Pr=0.71$, which can be explained by the higher thermal diffusivity at low Prandtl numbers. These observations support the model’s capability in capturing key flow features, even under conditions for which reference data are lacking.

\subsection{Distribution function of temperature fluctuations}
To gain a deeper understanding of the radial statistical asymmetry, detailed statistics of temperature fluctuations are computed for various distances from the cylindrical inner and outer wall by making use of the high-fidelity representation of an evolving turbulent state in ODT. The probability density function (PDF) of the collected normalized temperature fluctuations is then analyzed further. Two representative locations on both the inner and outer wall are selected: $r^+=5$, located in the viscous-conductive sublayer, and $r^+=100$, located in the turbulent-heat-flux-dominated region where turbulence is more significant. 

\begin{figure}
    \centering
    \hspace*{-0.05\textwidth}
    \includegraphics[width=1.1\textwidth]{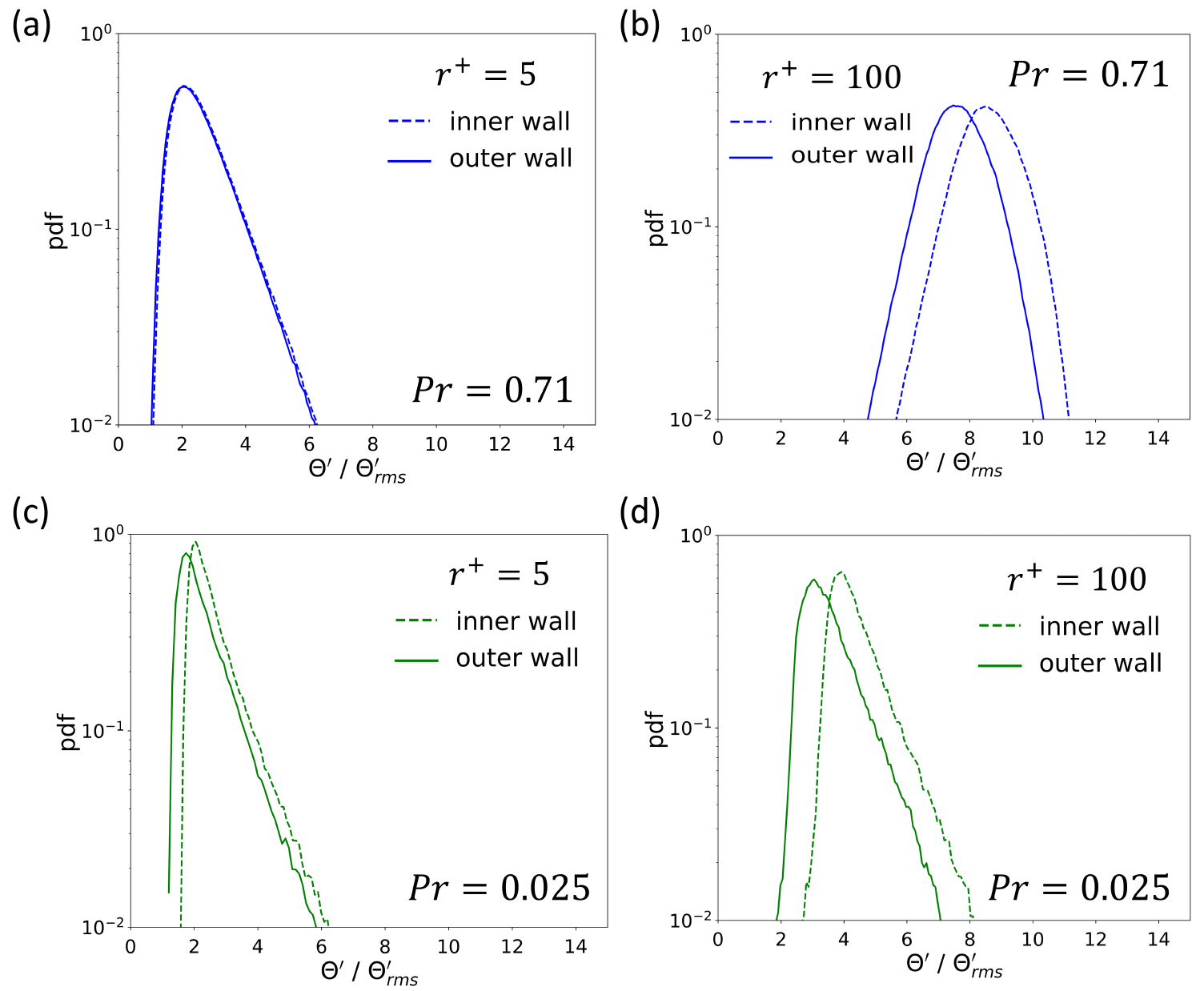}
    \caption{Probability density functions (PDFs) of the temperature fluctuations $\Theta^\prime$ normalized by the RMS temperature $\Theta^\prime_{\rm rms}$ at $r^+=5$ (in the conduction-dominated region) and $r^+=100$ (in the turbulent-heat-flux-dominated region) over the inner and outer cylinder walls for fixed radius ratio $\eta=0.1$ and bulk Reynolds number $Re_{D_{\rm h}}=10^5$.
    }
    \label{fig:PDF}
\end{figure}

Figure~\ref{fig:PDF} presents the PDFs of the normalized temperature fluctuations, $\Theta^\prime/\Theta^\prime_{\rm rms}$, for two different Prandtl numbers, $Pr=0.025$ and $Pr=0.71$, sampling at selected $r^+$ locations over the inner and outer wall. The radius ratio is fixed at $\eta=0.1$ and the bulk Reynolds number at $Re_{D_{\rm h}}=10^5$. The PDFs exhibit a unimodal distribution with noticeable skewness, rather than a symmetric Gaussian shape. For the low-Prandtl-number case $Pr=0.025$, the PDFs is narrower and exhibits a sharper cusp than the PDF for $Pr=0.71$, suggesting less extreme excursions from the mean state in more diffusive regimes.

At the near-wall location $r^+=5$, the of PDF of the temperature fluctuations at the inner and outer wall, respectively, are nearly collapsed onto each other for $Pr=0.71$, indicating a similar intensity of thermal fluctuations on both sides. By contrast, for $Pr=0.025$, radial asymmetry becomes more noticeable. In particular, the magnitude of $\Theta^\prime/\Theta^\prime_{\rm rms}$ is consistently higher on the inner wall, which is attributable to wall curvature. As the wall-normal distance increases to $r^+=100$, this asymmetry becomes more pronounced. For both $Pr$ investigated, the PDF over the cylindrical inner wall exhibits a similar peak height and spread as its outer wall counterpart, but with temperatures shifted by one standard deviation. In the $Pr=0.71$ case, the magnitude of temperature fluctuations shows a more substantial increase than in the $Pr=0.025$ case. The overall shape of the PDFs remains qualitatively similar between the inner and outer walls.

\section{Conclusions}\label{sec:conclusions}
Wall-curvature effects on scalar transfer in tubular heat exchangers are investigated for weakly heated turbulent annular pipe flow by variation of the radius ratio and the Reynolds number for a near-unity and a low Prandtl number. Predictive numerical simulations are performed using the One-Dimensional Turbulence (ODT) model as a standalone tool, extending previous work on momentum transfer in turbulent annular pipes~\cite{tsai2026}. The model is applied with full-scale resolution of the instantaneous flow and temperature profile across the annular gap, directly resolving the radial boundary layers on the inner and outer cylindrical walls without closure and closure modeling assumptions. Instead, a building-block-based approach is adopted in which the collective effect of spatial mapping events models turbulent fluxes in radial direction, reproducing boundary layer turbulence on the basis of mixing-length and attached-eddy phenomenology~\cite{klein2022,nath2026}. A previously calibrated ODT model setup for annular pipe flow is taken and validated here against available reference direct numerical simulation (DNS) data~\cite{bagheri2021} at a bulk Reynolds number of $Re_{D_{\rm h}}=1{,}77\times 10^4$ (based on the hydraulic diameter), molecular Prandtl number $Pr = 0.71$, and radius ratio $\eta = 0.1$.

The heat transfer scaling properties are investigated in terms of a global (bulk) and a local (gradient-based) Nusselt number. The latter is affected by wall curvature (radius ratio) for prescribed and constant wall heat fluxes. It is found that, even at high Reynolds numbers, curvature effects persist in the bulk Nusselt number, though they are not significant. Empirical correlations from~\cite{kays1980,sleicher1975} can approximate the trend and magnitude of the Nusselt number from $Re_{D_{\rm h}}=1{,}77\times 10^4$ to $Re_{D_{\rm h}}=10^6$, but they fail to capture the full behavior accurately due to a lack of consideration for the curvature effect. Our analysis of the local Nusselt number reveals that the inner wall Nusselt number $Nu_{\rm i}$ is more strongly influenced by curvature than the outer wall value $Nu_{\rm o}$, which is revealed by empirical correlation functions which have been obtained as fitting functions to the ODT results. The investigation of low-order and detailed statistics of temperature fluctuations, together with the application of boundary layer theory, points to a radial asymmetry of the flow and temperature fields. This asymmetry is geometrically induced and has nonlocal influence due to flux conservation properties that manifest themselves in a radial modification of the mixing length model, which in ODT is a consequence of radially modified mapping events~\cite{lignell2018}.  Theoretical analysis of the mean state and the turbulent fluxes reveals a physical compatibility of Reynolds-averaged ODT and Navier--Stokes balance equations, which allows for the capture of the dependence of the turbulent heat transfer on fluid properties (Prandtl number) and heat exchanger geometry (radius ratio). Across all cases investigated, the influence of curvature diminishes with increasing Reynolds number. Additionally, when the Prandtl number is decreased from $Pr=0.71$ to $0.025$, near-wall temperature gradients on both the inner and outer walls decrease, leading to a reduction in the local Nusselt number. Temperature fluctuations are also reduced as $Pr$ decreases, but they exhibit slightly stronger dependence on curvature compared to the higher Prandtl number case at the same Reynolds number. To complement the numerical results, we develop an improved thermal boundary layer analysis that accounts for curved-wall geometries, which is an extension of previous works~\cite{boersma2011,tsai2026}, addressing the limitations of traditional flat-wall models. This revised analysis is shown to remain valid up to $Re_{D_{\rm h}}=10^6$ investigated and performs reliably even under low Prandtl number conditions. The findings of this study offer a dimensionally reduced and dynamically modeling approach for the detailed simulation of tubular heat exchangers, applicable to variable and transient conditions. The heat transfer scaling correlations and parametric dependencies derived can be utilized directly in system models, improving the representation of the heat exchanger geometry.  

\section*{Acknowledgement}
This research is supported by the German Federal Government, the Federal Ministry of Research, Technology and Space and the State of Brandenburg within the framework of the joint project EIZ: Energy Innovation Center (project numbers 85056897 and 03SF0693A) with funds from the Structural Development Act (Strukturstärkungsgesetz) for coal-mining regions.










\bibliographystyle{elsarticle-num}

\bibliography{literature}

\end{document}